\documentclass[reprint,
superscriptaddress,
 amsmath,amssymb,
 aps,
prl,
]{revtex4-2}

\usepackage{graphicx}
\usepackage{dcolumn}
\usepackage{bm}
\usepackage{xcolor}
\usepackage[normalem]{ulem} 
\usepackage[hidelinks]{hyperref}

\begin{document}

\preprint{APS/123-QED}

\title{Particle-resolved study of the onset of turbulence}

\author{E. Joshi}
\email{eshita.joshi@dlr.de}
\affiliation{Institut f\"ur Materialphysik im Weltraum, Deutsches Zentrum f\"ur Luft- und Raumfahrt (DLR), 51147 K\"oln, Germany}
\author{M. H. Thoma}
\affiliation{I. Physikalisches Institut, Justus–Liebig
Universit\"at Gie{\ss}en, 35392 Gie{\ss}en, Germany}
\author{M. Schwabe}
\affiliation{Deutsches Zentrum f\"ur Luft- und Raumfahrt (DLR), Institut f\"ur Physik der Atmosph\"are, 82234 Oberpfaffenhofen, Germany}

\date{\today}

\begin{abstract}
The transition from laminar to turbulent flow is an immensely important topic that is still being studied. Here we show that complex plasmas, i.e., microparticles immersed in a low temperature plasma, make it possible to study the particle-resolved onset of turbulence under the influence of damping, a feat not possible with conventional systems. We performed three-dimensional (3D) molecular dynamics (MD) simulations of complex plasmas flowing past an obstacle and observed 3D turbulence in the wake and fore-wake region of this obstacle. We found that we could reliably trigger the onset of turbulence by changing key parameters such as the flow speed and particle charge, which can be controlled in experiments, and show that the transition to turbulence follows the conventional pathway involving the intermittent emergence of turbulent puffs. The power spectra for fully developed turbulence in our simulations followed the -5/3 power law of Kolmogorovian turbulence in both time and space. We demonstrate that turbulence in simulations with damping occurs after the formation of shock fronts, such as bow shocks and Mach cones. By reducing the strength of damping in the simulations, we could trigger a transition to turbulence in an undamped system. This work opens the pathway to detailed experimental and simulation studies of the onset of turbulence on the level of the carriers of the turbulent interactions, i.e., the microparticles.
\end{abstract}

\keywords{Onset of turbulence, complex plasmas, fluid dynamics, Kolmogorov turbulence, bow shocks, mach cones}
\maketitle

\begin{figure}
    \includegraphics[width=\linewidth]{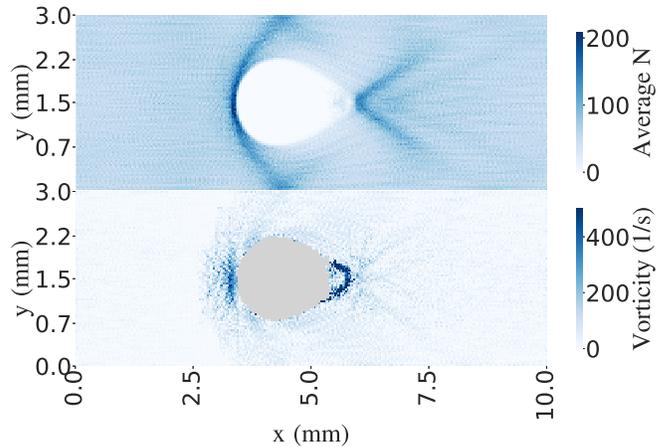}
    \caption{An example turbulent simulation with particles flowing from the left to the right seen in a slice of position $z = (1.9$ to $2.1)$~mm. (top) Number of particles, and (bottom) $z$ vorticity component are averaged in bins of size $0.03 \times 0.03 \times 0.2~\mathrm{mm}^3$ over 1~s. The empty pixels are coloured in grey to mark the obstacle and the cavity around it. Turbulence can be seen in the fore-wake region in front of the bow shock, and in the wake region near the Mach cones.}
    \label{fig:example_turb}
\end{figure}

One of the oldest unsolved problems in physics is the problem of the onset of turbulence. The transition of a fluid from smooth, deterministic, `laminar' sheets of flow to a chaotic, disordered, `turbulent' flow with vortices and eddies has been studied extensively in various systems such as water~\cite{Nazarenko2016, Falcon2022}, air~\cite{Busch1968}, as well as plasmas~\cite{Hasegawa1983, Goldman1984, Comisso2022, Marino2023}, and yet the ability to predict the onset of turbulence remains poor. A common way to trigger turbulence in experiments and simulations is to study flow past an obstacle~\cite{Tomboulides2000, Charan2018, Ortiz2019, Puthan2022}, as seen, for example, in Fig.~\ref{fig:example_turb}. Most investigations of flow past an obstacle focus on the wake region~\cite{Zhang1995, Wang2010, Dynnikova2016}, with little emphasis on the fore-wake region~\cite{Haakonsen2015, Tiwari2016}.

There is no generally agreed upon definition of turbulence~\cite{Groisman2000}, rather only signatures of dynamic behavior that are widely recognized to be associated with it~\cite{Mathieu2000}. Turbulent flows are~\cite{Nazarenko2011}:

\begin{enumerate}
    \itemsep0em 
    \item \textbf{Rotational.} Turbulence is accompanied by vortices and non-zero angular momentum. \label{item:rotational}
    \item \textbf{Chaotic.} Turbulence is characterized by irregularities in the flow in the form of rapid changes in velocity vectors, temperature, pressure, density, etc. The flow is disordered, and changes in the flow profile are intermittent, abrupt, and unpredictable.~\label{item:chaotic}
    \item \textbf{Diffusive.} The rate of mixing in fluids is higher when the flow is turbulent, especially in the direction perpendicular to the flow. \label{item:diffusive}
    \item \textbf{Resistive.} Turbulence creates drag. Flow resistance increases in turbulent flows compared to that of laminar flow. \label{item:resistive}
    \item \textbf{Self-similar.} Kolmogorov showed in 1941 that large vortices transfer energy into smaller vortices in an \textit{energy cascade} until the energy is dissipated through molecular diffusion and viscosity~\cite{Kolmogorov1941a, Kolmogorov1991}. The power spectra of energy cascades follow the power law $E(k) \propto k^{n}$, where $k$ is the wavenumber representing length scales. The equivalent power law $E(f) \propto f^{n}$ is true for the energy cascade in time, where $f$ represents frequency. For fully developed isotropic turbulence, n = $-5/3$~\cite{Frisch1995_book}.~\label{item:self-similar}
\end{enumerate}

The main parameter characterizing turbulent flow is the Reynolds number, $\Re = \mathrm{v}L/\nu$, where $\mathrm{v}$, $L$, and $\nu$ represent the characteristic flow speed, characteristic length scale, and kinematic viscosity, respectively. In this work, we use the bulk flow speed and the size of the cavity created by the obstacle for $\mathrm{v}$ and $L$, respectively. Generally, the transition from laminar flow to turbulent flow requires very high Reynolds numbers $\Re \geq O(1000)$~\cite{Hof2005, Smits2011}, however, low-$\Re$ turbulence has also been observed in many systems~\cite{Wang2014, Linkmann2019}. Studying low Reynolds number turbulence is important for fluids in biological systems and active matter, for instance, the flow of blood in vessels~\cite{Ghalichi1998}, or active swimmers such as bacteria~\cite{Peng2021}, as well as highly viscous polymer flows in the case of elastic turbulence~\cite{Steinberg2021}. 

One such system with very low Reynolds number values, $\Re \sim O(10)$, is a complex plasma. Complex plasmas are low-pressure and temperature plasmas with micrometer-sized particles embedded in them~\cite{Morfill2009}. These particles collect charges on their surfaces to become highly negatively charged, and as a result can become strongly coupled. Experiments are typically performed at pressures of the background gas of the order of (1-50)~Pa, which means that the microparticle dynamics are not over-damped as in colloids~\cite{Ivlev2012a}, but the Epstein damping force resulting from friction with the background gas is nevertheless an important force acting on the microparticles~\cite{Epstein1924}. Turbulence in complex plasmas has been primarily studied in the context of experiments on waves~\cite{Pramanik2003,Tsai2012,Tsai2020,Zhdanov2015, Schwabe2018, Hu2022} and simulations of fully developed turbulence~\cite{Schwabe2014}, or 2D turbulence~\cite{Gogia2020, Kostadinova2021}. Studying turbulence in complex plasmas has the advantage that each microparticle is large enough to be imaged directly, allowing for particle-resolved studies~\cite{Bajaj2023}. This will help understand the onset of turbulence in more detail, and to eventually develop methods to better control it. 

In this Letter, we demonstrate how complex plasmas can be used to study the onset of turbulence in detail. For this, we performed 3D molecular dynamics (MD) simulations of subsonic and supersonic complex plasma flow past an obstacle with turbulence in the wake and fore-wake regions. We will first present an example of a turbulent simulation and show that it is consistent with the hallmarks of turbulence listed above. We will then discuss the transition of laminar flow to turbulent flow as well as transition from turbulence in a damped to an undamped system as observed in our simulations. We found that we can reliably trigger the onset of turbulence by changing the flow speed and particle charge.

\begin{figure}
    \centering
    \textbf{a})\includegraphics[width=\linewidth]{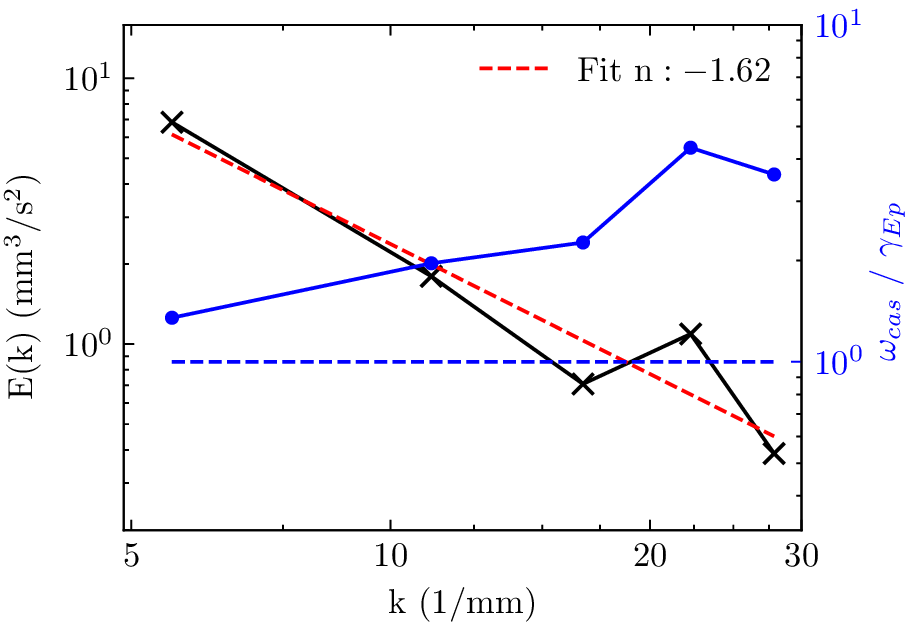}\\
    \textbf{b})\includegraphics[width=\linewidth]{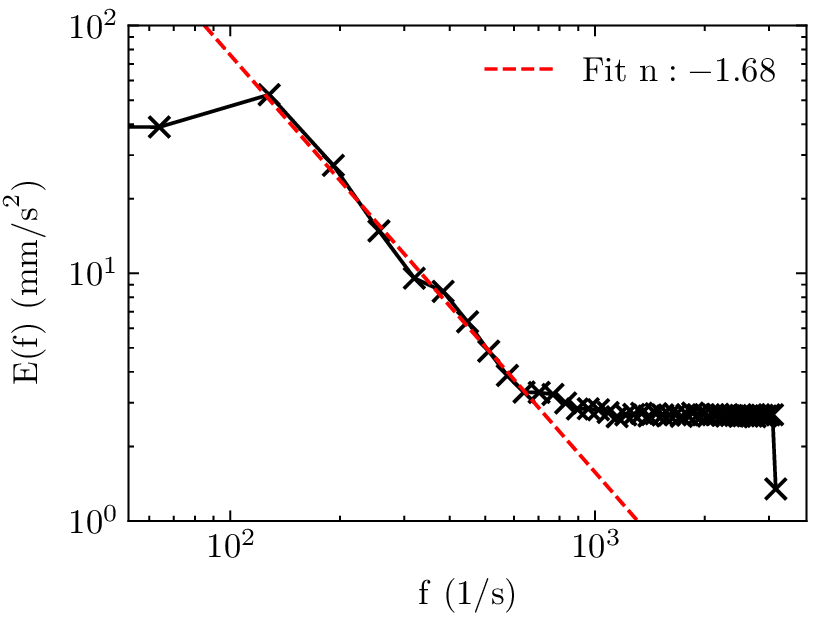}
    \caption{Power spectra in \textbf{a}) space and \textbf{b}) time corresponding to the simulation in Fig.~\ref{fig:example_turb}. Power spectra are shown from the interparticle distance to the size of the turbulent region in space, and till the Nyquist frequency of 5000~Hz in time. Both subplots show approximately the $n = -5/3$ power law characteristic of turbulent flows. The cascade rate $\omega_\textrm{cas}$ is greater than the damping rate $\gamma_\textrm{Ep}$, showing that the system was in the regime where turbulence could develop. The red dashed lines show the linear fit through the data, whereas the horizontal blue dashed line shows where $\omega_\textrm{cas} = \gamma_\textrm{Ep}$.}
    \label{fig:power_spectra}
\end{figure}

We used the open source software `Large-scale Atomic/Molecular Massively Parallel Simulator' (LAMMPS) \cite{Plimpton1995, Thompson2022} to generate the simulations. We typically used a time step of 0.1~ms and a data acquisition rate of 1000~fps, and a simulation box with dimensions $10 \times 3 \times 3~\mathrm{mm}^{3}$ with periodic boundary conditions. The data acquisition rate was changed to 10,000~fps to generate the power spectra in time given in Fig.~\ref{fig:power_spectra}. We simulated microparticles as point charges with an interparticle distance of $150$~$\mu$m with a Yukawa interparticle potential. The particle charge used was $q=-3481$~e (elementary charges) and the Debye length was $\lambda_\mathrm{D} = 100$~$\mu$m~\cite{Schwabe2014}. The mass of the point charges was set to 3~$\times 10^{-14}$~kg, mimicking microparticles with mass density $\rho = 1510$~kg/m$^3$ and diameter $d = 3.38~\mu$m. We used a Langevin dynamics thermostat~\cite{Schneider1978} with an Epstein damping coefficient~\cite{Epstein1924} corresponding to argon gas at a pressure of $10$~Pa. The main forces acting on the particles in the simulation are, therefore, the interparticle force given by the Yukawa potential, the Langevin force due to the thermostat, and the Epstein damping due to the friction with the background gas.

We included the obstacle using a point charge of $-3\times10^{7}$~e with a fixed position at the coordinates $(x, y, z) = (4.5, 1.5, 2.0)$~mm. As observed in experiments~\cite{Meyer2014, Jaiswal2018, Bailung2020}, this high negative charge repelled the microparticles and thus resulted in a cavity around the point charge. In simulations in which the particles at a temperature of 2000~K were stationary, the cavity was spherical with a diameter of $1.4$~mm. Please note that due to the periodic boundary conditions used in all dimensions, the simulation actually consisted of an infinite grid of obstacles. Increasing the size of the simulation box to ensure that the disturbed flow from one obstacle did not interfere with that around another obstacle showed no changes to our results.

We created a flow of particles past the obstacle by applying a constant force $F_\mathrm{ext} = 10^{-11} - 10^{-13}$~N in $x$-direction to the microparticles, depending on the desired bulk flow speed. An external force of $F_\mathrm{ext} = 10^{-13}$~N could be produced in experiments by an electric field with a magnitude of 180~V/m. Due to the damping force, this resulted in a steady flow from the left to the right of the simulation box with periodic boundary conditions. The simulation box was large enough along $x$ for the particle motion to return to laminar before reaching the boundaries of the box. The main parameter used to characterize the flow at speed $\mathrm{v}$ is the Mach number, $\mathcal{M} = \mathrm{v}/C_{s}$, where $C_{s} = \sqrt{Z_{d}k_{B}T_{d}/m_{d}}$  is the speed of sound in the system, with $m_{d}$ as the microparticle mass and $Z_{d} = q/e$. For supersonic velocity $\mathcal{M} > 1$.

\begin{figure}
    \includegraphics[width=\linewidth]{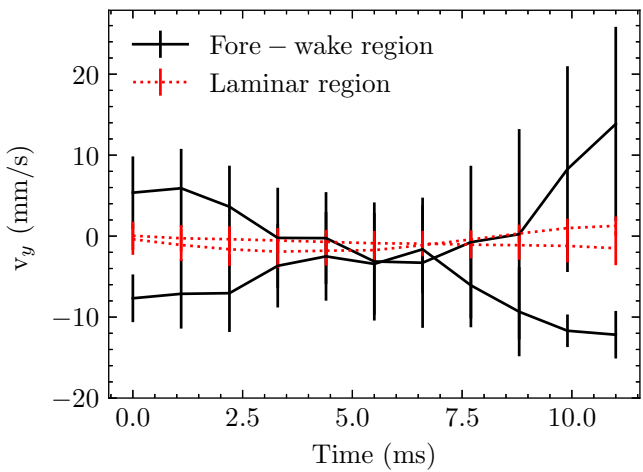} 
    \caption{Plot of the mean and standard error of the flow velocity $\mathrm{v}_{y}$ (perpendicular to the bulk flow direction) of two sets of five randomly chosen particles within the laminar (red dotted lines) and fore-wake (black solid lines) region. In the laminar region, particle velocities $\mathrm{v}_{y}$ have little fluctuations over time implying a regular and ordered flow with almost no fluid mixing. On the other hand, $\mathrm{v}_{y}$ in the fore-wake region shows large random fluctuations on very short timescales, characteristic of intermittency. This indicates an irregular and disordered flow state with increased fluid mixing.}
\label{fig:mixingc}
\end{figure}

Previous studies of supersonic flow have reported turbulence as well as Mach cones in the wake of an obstacle~\cite{Samsonov1999, Melzer2000, Morfill2004, Jiang2009, Charan2016}. However, there have been few investigations of fore-wake turbulence or on the importance of shock fronts in triggering the onset of turbulence in damped systems. Hence, we will be focusing on fore-wake turbulence in this work. An example of wake and fore-wake turbulence in our simulation is provided in Fig.~\ref{fig:example_turb}, with an accompanying movie in the Supplemental Material~\cite{Supplement}. This is a 2D slice of the simulation box which shows the spatial distribution of vorticity $\mathbf{\omega}_{z} = \frac{\partial \mathrm{v}_y}{\partial x} - \frac{\partial \mathrm{v}_x}{\partial y}$ around the obstacle averaged over one second of simulation time with a bin size of $0.03 \times 0.03 \times 0.2~\mathrm{mm}^{3}$. The vorticity per bin was calculated by averaging the difference in $\mathrm{v}_{x}$ and $\mathrm{v}_{y}$ across neighboring horizontal and vertical bins respectively. Shock fronts such as Mach cones and a bow shock can be seen in Fig.~\ref{fig:example_turb}(top). Fig.~\ref{fig:example_turb}(bottom) shows that  vortices were generated in two main regions. The first is the wake of the obstacle, where the particles deflected by the obstacle flowed into the Mach cone. The second is the fore-wake region of the obstacle, where the incoming particles flowed into the bow shock. In both cases, vorticies were induced when particles flowed directly into a shock front. This observation is similar to the experimental observation of turbulence in Ref.~\cite{Bajaj2023}, where the flow of particles directly into a dust-acoustic wavefront induced turbulence. The existence of vortices in the wake and fore-wake regions of the simulation satisfies Condition~\ref{item:rotational} of the hallmarks of turbulence.

In the following, we study the flow in two regions in detail, a laminar region with position 0 $\leq x \leq$ 1~mm, 1 $\leq y \leq$ 2~mm, 1 $\leq z \leq$ 2~mm and a turbulent region corresponding to the fore-wake with position 2.5 $\leq x \leq$ 3.5~mm, 0.8 $\leq y \leq$ 2.2~mm, and 0.8 $\leq z \leq$ 2.2~mm. The evolution of the particle velocities perpendicular to the flow direction is given in Fig.~\ref{fig:mixingc}. The particle motion in the laminar flow was smooth, ordered, and predictable, with almost no fluid mixing perpendicular to the flow direction. In contrast, the particle motion in the fore-wake was unpredictable and disordered, with irregularities in the flow and abrupt changes in the velocity. This is an example of intermittency in fully developed turbulence where the kinetic energy (and hence velocity) undergoes abrupt changes in a short time~\cite{Frisch1995_book, Carter2006}. There was also increased mixing of the fluid in the turbulent region, especially in the direction perpendicular to the flow, as seen by the increased magnitude of $\mathrm{v}_{y}$. This satisfies Conditions~\ref{item:chaotic} and \ref{item:diffusive} listed above.

The driving force causing the particles to flow from the left to the right of the simulation box was applied uniformly. Plots of the normalized kinetic energy, (ke~-~$\mu_{\mathrm{ke}}$)/$\sigma_{\mathrm{ke}}$, of the particles in the laminar and chaotic regions of the fore-wake, where $\mu_{\mathrm{ke}}$ and $\sigma_{\mathrm{ke}}$ are the mean and standard deviation of the particle kinetic energy at a given time, can be found in the Supplemental Material~\cite{Supplement}. These plots show that the kinetic energy in the chaotic region was significantly lower than that of the laminar flow for the same driving force. This indicates an increased flow resistance in the fore-wake region, satisfying Condition~\ref{item:resistive} of the hallmarks of turbulence.

To separate the kinetic energy due to the driving force and that due to the turbulent interparticle interactions, we define \textit{turbulent velocity} \textbf{v}$'$ and \textit{turbulent kinetic energy} TKE as
\begin{equation}
    \textbf{{v}}' = \textbf{{v}} - \overline{\textbf{v}} \quad\text{and}\quad
    \mathrm{TKE} = \frac{1}{2}m\mathrm{v}'^{2},
\end{equation}
where $\overline{\textbf{v}}$ is the local Reynolds-averaged flow velocity calculated with 3D bins of size $(250~\mu \mathrm{m})^3$ averaged over one second, and $m$ denotes the particle mass. We used $\textbf{v}'$ to calculate the power spectra in space, and $\textbf{v}$ to calculate the spectra in time.

The transfer of turbulent kinetic energy from large scales to smaller scales in 3D turbulence follows a power law distribution of $E(k) \propto k^n$ with n=$-5/3$, as first described by Kolmogorov~\cite{Kolmogorov1941a, Kolmogorov1991}. The one-dimensional power spectra in $x$, $y$, and $z$ can be found in the Supplemental Material~\cite{Supplement}. They all show the n=$-5/3$ power law, showing that turbulence is three-dimensional and isotropic in our system~\cite{Frisch1995_book}. The power spectra in time and 3D space are given in Fig.~\ref{fig:power_spectra}, which also reflect the n = $-5/3$ power law, further satisfying Condition~\ref{item:self-similar} of the signatures of turbulence. The power spectra were calculated using the Wiener-Khintchine theorem~\cite{Wiener1930, Khintchine1934}, which states that the power spectrum is the Fourier transform of the velocity correlation function. This is given by the relation
\begin{align}
    E(f) = \frac{1}{2\pi} \int_{-\infty}^{+\infty} e^{ifs} \Gamma(s) ds, \quad\text{and}\quad \\
    E(k) = \frac{1}{2\pi} \int_{-\infty}^{+\infty} e^{ikx} \left<\mathrm{v}(0)\mathrm{v}(x)\right> dx,
\end{align}
where $\Gamma(s) = \left<\mathrm{v}(t)\mathrm{v}(t')\right>$ is the velocity correlation function averaged over all particles, and $s = |t-t'|$. The power spectra follow the n = $-5/3$ law from f~$\sim$~110~Hz to f~$\sim$~660~Hz, and from spatial scales of roughly 225~$\mu$m to 1.14~mm. These values are consistent with the estimated Kolmogorov microscales~\cite{Landahl1992, Frisch1995_book}. The Kolmogorov dissipation length ($\eta \sim 230~\mu$m) and time ($1/\tau_n \sim 80$~Hz) were calculated using the mean rate of kinetic energy transfer, $\varepsilon$, and the local viscosity, $\nu$. The local viscosity was estimated using the average interparticle distance as the mean-free path, whereas $\varepsilon$ was determined numerically using $\frac{1}{2} \frac{\partial <\mathrm{v}^{2}>}{\partial t}$.
It can also be seen in Fig.~\ref{fig:power_spectra} that the rate of energy cascade, $\omega_{cas} = k\sqrt{2kE(k)}$, is greater than the rate of damping, $\gamma_{Ep}$ = 36~s$^{-1}$, given by the Epstein damping coefficient \cite{Epstein1924}, for all considered wave numbers. This shows that our system was in the regime where vortices could transfer energy from larger length scales to smaller scales without dissipation until the cascade reached interparticle distances \cite{Bajaj2023}. The five conditions representing the hallmarks of turbulence listed above, plus one specific for damped systems, are thus shown to be satisfied.

\begin{figure}
    \begin{center}
    \includegraphics[height=60mm]{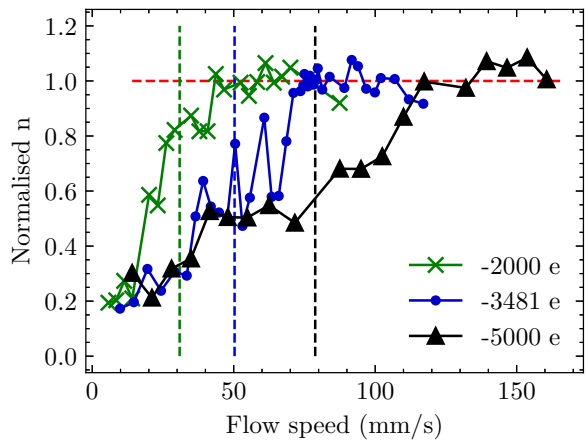}
    \caption{Dependence of the power spectrum slopes normalized by $-5/3$,  obtained via a linear fit in the region 100~Hz~$<$~f~$<$~700~Hz of the power spectra in time, on the flow speed and the particle charge. Vertical lines indicate the speed of sound for the corresponding particle charge defined as $C_{s} = \sqrt{Z_{d}k_{B}T_{d}/m_{d}}$ .}
    \label{fig:MRe}
    \end{center}
\end{figure}

In the following, we will discuss a series of simulations studying the onset of turbulence. The onset of this turbulent flow in the simulations was triggered by changing one of two main parameters: the flow speed and the particle charge. Changing the flow speed changed $\mathcal{M}$ and $\Re$. Changing the particle charge affected viscosity, and hence $\Re$, but also the speed of sound $C_{s} = \sqrt{Z_{d}k_{B}T_{d}/m_{d}}$ and hence $\mathcal{M}$. Increasing $\mathcal{M}$ led to the formation of shock fronts, such as Mach cones and bow shocks. These shock fronts were crucial for triggering the onset of turbulence in our simulations. In the simulations with damping, no turbulence was observed in the fore-wake without the presence of shock fronts, even with particles flowing around the obstacle at supersonic speeds. In the fore-wake region, the laminar flow of particles flowing directly perpendicular into the bow shock triggered the flow to become turbulent. Similarly, particles accelerated by the obstacle from above and below that flowed directly into the Mach cone in the wake became turbulent. The significance of the presence of bow shocks and Mach cones can be seen in Fig.~\ref{fig:MRe}, which plots the slope of the power spectra (normalised by -5/3) as a function of average flow speed for three different particle charges for otherwise identical conditions. The vertical dashed lines mark the speed where the flow becomes supersonic, with colors corresponding to the three simulated cases of particle charge. It can be seen clearly that the normalized slope reaches a value of 1 only after the flow becomes supersonic.

The shape of the plots can be explained by the following. For a given supersonic flow speed, the onset of turbulence could be triggered by decreasing the viscosity, and hence increasing $\Re$. This was achieved by decreasing the magnitude of the particle charge and keeping all other parameters constant. If $\Re$ is high enough for turbulence to develop, for instance for a particle charge of -2000~e, $\Re \sim O(10)$, then the flow becomes turbulent as soon as $\mathcal{M} > 1$. If $\Re$ is low, for instance for a particle charge of -5000~e, $\Re \sim O(1)$, then upon reaching $\mathcal{M} > 1$ no turbulence develops. $\Re$ increases linearly with increasing flow speed until it is high enough for turbulence to develop. For higher particle charges, $\Re$ is lower and thus the system requires higher $\mathcal{M}$ to become turbulent.

\begin{figure}
    \centering\includegraphics[width=\linewidth]{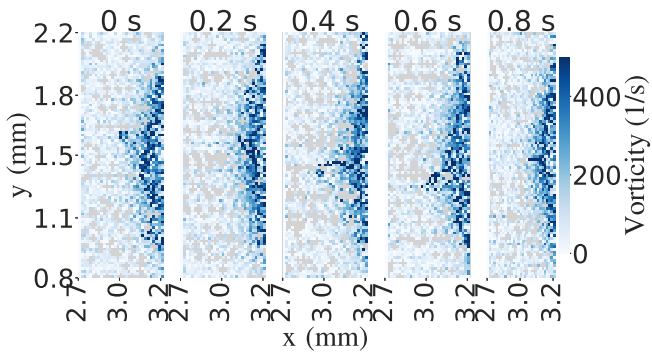}
    \caption{Vorticity heatmap showing intermittent switching between laminar and turbulent regions as transient turbulent puffs form and decay over space and time.}
    \label{fig:intermittent}
\end{figure}

For supersonic flow velocities during the transition to turbulence, intermittent switching between laminar and turbulent regions was also observed. This is shown in Fig.~\ref{fig:intermittent}. The flow in the fore-wake region was observed to switch from laminar to turbulent and back to laminar as transient turbulent puffs were created over time. This is characteristic of intermittent turbulence occurring during the transition to turbulence~\cite{Eckhardt2007, Gogia2020}. Animations of vorticity over time showing intermittent turbulence can be found in the Supplemental Material~\cite{Supplement}.

The result that the formation of shock fronts is necessary for turbulence to develop is due to the presence of damping in typical conditions for complex plasma experiments that informed our simulations. To confirm this, we performed 3D simulations in a practically undamped system with supersonic flow, with a pressure corresponding to $10^{-5}$~Pa, and repeated the same analysis steps as before. The criteria for turbulence listed above were satisfied, and we observed turbulence in the undamped simulations even without the presence of shock fronts. The results of this analysis can be found in the Supplemental Material~\cite{Supplement}.

To summarize, we report the results of 3D molecular dynamics simulations of a complex plasma with turbulent flow in the wake and fore-wake region of a supersonic obstacle. We confirmed that the flow indeed was turbulent by going through a checklist of the hallmarks of turbulence. We studied the onset of turbulence in multiple simulations and found that we could reliably trigger the onset of turbulence by changing parameters such as flow speed and particle charge. Turbulence was observed in the simulations with damping if and only if there was a flow of particles directly towards a shock front. We observed no turbulence in the absence of bow shocks and Mach cones. We demonstrated a dependence of the power spectra of energy cascade on $\mathcal{M}$ and $\Re$, by studying their evolution with increasing flow speed and particle charge. Intermittency was observed both during the onset of turbulence (in the form of transient turbulent puffs) and in fully developed turbulence (in the form of rapid changes to velocity vectors). The flow speeds, particle charges and pressure regime used in this study are obtainable with typical complex plasma experimental setups. Therefore, we expect that our study can be used to inform future experiments and open up new research avenues for detailed particle-resolved studies of the onset of turbulence using complex plasmas.

\acknowledgments
  The authors gratefully acknowledge funding of this work in the framework of the Nachwuchsgruppenprogramm im DLR-Geschäftsbereich Raumfahrt. We would like to thank Volodymyr Nosenko, Prapti Bajaj and Alexei Ivlev for helpful discussions and their valuable feedback, and Hubertus Thomas and Janis Klamt for careful reading of the manuscript and for helpful comments.

\bibliographystyle{apsrev4-2}
\bibliography{turbulence}

\end{document}